\documentstyle[12pt,amssymb]{article}
\def\hybrid{\topmargin 0pt      \oddsidemargin 0pt
        \headheight 0pt \headsep 0pt
        \textwidth 16.5cm
        \textheight 23cm
        \voffset=-1cm
        \hoffset=0.4cm
        \marginparwidth 0.0in
        \parskip 5pt plus 1pt   \jot = 1.5ex}
\catcode`\@=11
\def\marginnote#1{}

\newcount\hour
\newcount\minute
\newtoks\amorpm
\hour=\time\divide\hour by60
\minute=\time{\multiply\hour by60 \global\advance\minute by-\hour}
\edef\standardtime{{\ifnum\hour<12 \global\amorpm={am}%
        \else\global\amorpm={pm}\advance\hour by-12 \fi
        \ifnum\hour=0 \hour=12 \fi
      \number\hour:\ifnum\minute<10 0\fi\number\minute\the\amorpm}}
\edef\militarytime{\number\hour:\ifnum\minute<10 0\fi\number\minute}

\def\draftlabel#1{{\@bsphack\if@filesw {\let\thepage\relax
   \xdef\@gtempa{\write\@auxout{\string
      \newlabel{#1}{{\@currentlabel}{\thepage}}}}}\@gtempa
   \if@nobreak \ifvmode\nobreak\fi\fi\fi\@esphack}
        \gdef\@eqnlabel{#1}}
\def\@eqnlabel{}
\def\@vacuum{}
\def\draftmarginnote#1{\marginpar{\raggedright\scriptsize\tt#1}}

\def\draft{\oddsidemargin -0.1truein
        \def\@oddfoot{\sl preliminary draft \hfil
        \rm\thepage\hfil\sl\today\quad\militarytime}
        \let\@evenfoot\@oddfoot \overfullrule 3pt
        \let\label=\draftlabel
        \let\marginnote=\draftmarginnote
\def\@eqnnum{{\rm (\theequation)}
\rlap{\kern\marginparsep\tt\@eqnlabel}%
\global\let\@eqnlabel\@vacuum}  }

\newcommand{\RR}{{\mathbb{R}}}
\newcommand{\CC}{{\mathbb{C}}}
\newcommand{\NN}{{\mathbb{N}}}
\newcommand{\ZZ}{{\mathbb{Z}}}            

\newfont{\Bbbb}{msbm7 scaled 1\@ptsize00}
\newcommand{\zs}{\raise-1pt\hbox{$\mbox{\Bbbb Z}$}}
\newcommand{\rs}{\hbox{$\mbox{\Bbbb R}$}}

\font\teneufm=cmmib10 scaled 1\@ptsize00
\font\seveneufm=cmmib7 scaled 1\@ptsize00
\font\fiveeufm=cmmib5  
\def\bfit#1{{\textfont1=\teneufm\scriptfont1=\seveneufm
\scriptscriptfont1=\fiveeufm
\mathchoice{\hbox{$\mathsurround=0pt\displaystyle#1$}}
{\mathsurround=0pt\hbox{$\textstyle#1$}}
{\hbox{$\mathsurround=0pt\scriptstyle#1$}}
{\hbox{$\mathsurround=0pt\scriptscriptstyle#1$}}}}
\font\sevenmsa=msam6 
\newfam\msafam
\textfont\msafam=\sevenmsa
\def\hexnumber@#1{\ifnum#1<10 \number#1\else
\ifnum#1=10 A\else\ifnum#1=11 B\else\ifnum#1=12 C\else
\ifnum#1=13 D\else\ifnum#1=14 E\else\ifnum#1=15 F\fi\fi\fi\fi\fi\fi\fi}
\def\msa@{\hexnumber@\msafam}
\def\llcorner{\delimiter"4\msa@78\msa@78 }
\def\lrcorner{\delimiter"5\msa@79\msa@79 }
\mathchardef\blacktriangleright="3\msa@49
\mathchardef\blacktriangleleft="3\msa@4A
\mathchardef\trianglerighteq="3\msa@44
\mathchardef\trianglelefteq="3\msa@45
\font\tenmsb=msbm10 scaled 1\@ptsize00
\newfam\msbfam
\textfont\msbfam=\tenmsb
\scriptfont\msbfam=\tenmsb
\def\msb@{\hexnumber@\msbfam}
\mathchardef\varkappa="0\msb@7B

\newdimen\linethick  \linethick=0.4pt
\newdimen\hboxitspace    \hboxitspace=5pt
\newdimen\vboxitspace    \vboxitspace=5pt

\def\fr#1{%
\beq\new
\vcenter{
\hrule height\linethick
           \hbox{\vrule width\linethick
                 \kern\hboxitspace
                 \vbox{\kern\vboxitspace
                       \hbox{$\begin{array}{c}\displaystyle#1
          \end{array}$}%
                       \kern\vboxitspace}%
                 \kern\hboxitspace
                 \vrule width\linethick}%
           \hrule height\linethick}%
\eeq}

\newdimen\Squaresize \Squaresize=14pt
\newdimen\Thickness \Thickness=0.5pt

\def\Square#1{\hbox{\vrule width \Thickness
   \vbox to \Squaresize{\hrule height \Thickness\vss
      \hbox to \Squaresize{\hss#1\hss}
   \vss\hrule height\Thickness}
\unskip\vrule width \Thickness}
\kern-\Thickness}

\def\Vsquare#1{\vbox{\Square{$#1$}}\kern-\Thickness}

\def\numberbysection{\@addtoreset{equation}{section}
        \def\theequation{\thesection.\arabic{equation}}}
\numberbysection

\renewcommand{\theequation}{\thesection.\arabic{equation}}
\newcommand{\l@qq}[2]{\addvspace{2em}
 \hbox to\textwidth{\hspace{1em}\bf #1 \dotfill #2}}

\newcounter{app}

\def\app{\setcounter{equation}{0}
\def\theequation{\Alph{app}.\arabic{equation}}\par
   \addvspace{4ex}
   \@afterindentfalse
  \secdef\@app\@dapp}
\newcommand\@app{\@startsection {app}{1}{0ex}%
                             {-3.5ex \@plus -1ex \@minus -.2ex}%
                                   {2.3ex \@plus.2ex}%
                                   {\normalfont\Large\bf}}

\def\@dapp#1{%
{\parindent \z@ \raggedright  \bf #1}\par\nobreak}
\def\l@app#1#2{\ifnum \c@tocdepth >\z@
    \addpenalty\@secpenalty
    \addvspace{1.0em \@plus\p@}%
    \setlength\@tempdima{2.5em}%
    \begingroup
      \parindent \z@ \rightskip \@pnumwidth
      \parfillskip -\@pnumwidth
      \leavevmode \bfseries
      \advance\leftskip\@tempdima
      \hskip -\leftskip
      #1\nobreak\hfil \nobreak\hb@xt@\@pnumwidth{\hss #2}\par
    \endgroup\fi}
\newcounter{sapp}[app]

\def\sapp{\def\theequation{\Alph{app}.\arabic{equation}}\par
   \@afterindentfalse
  \secdef\@sapp\@dsapp}
\newcommand\@sapp{\@startsection{sapp}{2}{\z@}%
                         {-3.25ex\@plus -1ex \@minus -.2ex}%
                           {1.5ex \@plus .2ex}%
                              {\normalfont\large\bfseries}}

\def\@dsapp#1{%
{\parindent \z@ \raggedright  \bf #1}\par\nobreak}
\newcommand{\l@sapp}{\@dottedtocline{2}{1.5em}{3em}}

\def\titlepage{\@restonecolfalse\if@twocolumn\@restonecoltrue\onecolumn
     \else \newpage \fi \thispagestyle{empty}\c@page\z@
        \def\thefootnote{\fnsymbol{footnote}} }

\def\endtitlepage{\if@restonecol\twocolumn \else  \fi
        \def\thefootnote{\arabic{footnote}}
        \setcounter{footnote}{0}}  
\relax

\hybrid

\newtoks\@stequation

\def\subequations{\refstepcounter{equation}%
  \edef\@savedequation{\the\c@equation}%
  \@stequation=\expandafter{\theequation}
  \edef\@savedtheequation{\the\@stequation}
  \edef\oldtheequation{\theequation}%
  \setcounter{equation}{0}%
  \def\theequation{\oldtheequation\alph{equation}}}

\def\endsubequations{%
  \setcounter{equation}{\@savedequation}%
  \@stequation=\expandafter{\@savedtheequation}%
  \edef\theequation{\the\@stequation}%
  \global\@ignoretrue}

\makeatletter
\newdimen\normalarrayskip            
\newdimen\minarrayskip               
\normalarrayskip\baselineskip
\minarrayskip\jot
\newif\ifold             \oldtrue            \def\new{\oldfalse}
\def\arraymode{\ifold\relax\else\displaystyle\fi}
\def\eqnumphantom{\phantom{(\theequation)}} 
\def\@arrayskip{\ifold\baselineskip\z@\lineskip\z@
     \else
     \baselineskip\minarrayskip\lineskip1\baselineskip\fi}

\def\@arrayclassz{\ifcase \@lastchclass \@acolampacol \or
\@ampacol \or \or \or \@addamp \or
   \@acolampacol \or \@firstampfalse \@acol \fi
\edef\@preamble{\@preamble
  \ifcase \@chnum
     \hfil$\relax\arraymode\@sharp$\hfil
     \or $\relax\arraymode\@sharp$\hfil
     \or \hfil$\relax\arraymode\@sharp$\fi}}

\def\@array[#1]#2{\setbox\@arstrutbox=\hbox{\vrule
     height\arraystretch \ht\strutbox
     depth\arraystretch \dp\strutbox
width\z@}\@mkpream{#2}\edef\@preamble{\halign \noexpand\@halignto
\bgroup \tabskip\z@ \@arstrut \@preamble \tabskip\z@ \cr}%
\let\@startpbox\@@startpbox \let\@endpbox\@@endpbox
  \if #1t\vtop \else \if#1b\vbox \else \vcenter \fi\fi
  \bgroup \let\par\relax
  \let\@sharp##\let\protect\relax
  \@arrayskip\@preamble}
\def\eqnarray{\stepcounter{equation}%
              \let\@currentlabel=\theequation
              \global\@eqnswtrue
              \global\@eqcnt\z@
              \tabskip\@centering              
              \let\\=\@eqncr
              $$%
            \halign to \displaywidth  \bgroup
             \eqnumphantom \@eqnsel
      \hskip\@centering                               
    $\displaystyle  \tabskip\z@ {##}$%
    &\global\@eqcnt\@ne \hskip 2\arraycolsep
         $ \displaystyle  \arraymode{##}$\hfil
    &\global\@eqcnt\tw@ \hskip 2\arraycolsep
         $\displaystyle\tabskip\z@{##}$\hfil
         \tabskip\@centering
    &{##}\tabskip\z@\cr}
\makeatother
\newtheorem{th}{Theorem}[section]

\newtheorem{lem}{Lemma}[section]

\newtheorem{rem}{Remark}[section]

\def\bea{\begin{eqnarray}}
\def\eea{\end{eqnarray}}

\def\beq{\begin{equation}}
\def\eeq{\end{equation}}
\def\be{\beq\new\begin{array}{c}}  
\def\ee{\end{array}\eeq}           
\def\bse{\begin{subequations}}                
\def\ese{\end{subequations}}

\def\square{\hfill{\vrule height6pt width6pt            
depth1pt} \break \vspace{.01cm}}                        

\def\bgamma{{\bfit\gamma}}

\def\blambda{{\bfit\lambda}}
\def\brho{{\bfit\rho}}

\def\la{\lambda}
\def\e{\epsilon}
\def\<{\langle}
\def\>{\rangle}
\def\ov{\overline}
\def\wt{\widetilde}
\def\wh{\widehat}
\def\N{\scriptscriptstyle N}
\begin{document}

\begin{titlepage}

\begin{center}
\hfill ITEP-TH-34/00\\

\phantom.
\bigskip\bigskip\bigskip\bigskip\bigskip\bigskip
{\Large\bf Integral representations for the eigenfunctions

\bigskip\noindent
of quantum open and periodic Toda chains

\bigskip\noindent
from QISM formalism\footnote{Talk given at the Workshop on MMRD 2000,
12-15 April 2000, Leeds} }\\

\bigskip \bigskip
{\large S. Kharchev\footnote{E-mail:  kharchev@vitep5.itep.ru}, D.
Lebedev\footnote{E-mail:  dlebedev@vitep5.itep.ru}}\\ \medskip {\it
Institute
of Theoretical \& Experimental Physics\\ 117259 Moscow, Russia}\\
\end{center}

\bigskip
\bigskip
\bigskip
\begin{abstract}
\noindent
The integral representations for the eigenfunctions of $N$ particle
quantum open and periodic Toda chains are constructed in
the framework of Quantum Inverse Scattering Method (QISM).
Both periodic and open $N$-particle solutions have essentially
the same structure being written as a generalized Fourier transform
over the eigenfunctions of the $N-1$ particle open Toda chain with
the kernels satisfying to the Baxter equations of the second and first
order respectively. In the latter case this leads to
recurrent relations which result to representation of the Mellin-Barnes
type for solutions of an open chain. As byproduct, we obtain the
Gindikin-Karpelevich formula for the Harish-Chandra function in the case
of $GL(N,\RR)$ group.
\end{abstract}

\end{titlepage}
\clearpage \newpage

\setcounter{page}1
\footnotesize
\normalsize

\section{Introduction}
This report is devoted to well-known quantum mechanical problem to
find the simultaneous eigenfunction of commuting set of
Hamiltonians for the periodic Toda chain. The first important step in
this direction has been done by Gutzwiller \cite{Gu} who solved the
problem for the particular cases $N=2,3$ and 4 particles and found
such important phenomena as quantization of spectrum and separation of
multidimensional Baxter equation into the product of one dimension ones.
In fact, he performed the quantization of periodic Toda chain in
terms of separated variables introduced by Flaschka and McLaughlin
\cite{FM}. Next important step was taken by Sklyanin \cite{Skl1} who
constructed $R$-matrix formalism for both classical and quantum cases
Toda chains and introduced the algebraic method of separation variables
for an arbitrary number of particles. His approach drastically simplifies
the derivation of the Baxter equation and works for wide spectrum of
integrable models \cite{Skl2}.

Our method to solve the spectral problem consists of analytical
re-interpretation of Sklyanin's algebraic ideas which allows to find the
integral representation for the eigenfunctions of the periodic Toda chain
as a kind of generalized Fourier transform with the eigenfunctions for
the open Toda chain \cite {KL1}. In turn, this method can be treated as a
natural generalization of original Gutzwiller's approach. The
explicit solution for the eigenfunctions of the open Toda chain plays
a key role in this construction.

It has been discovered by Kostant \cite{Konst} that
the commuting set of Hamiltonians of an open Toda chain coincides
with the Whittaker model of the center of universal enveloping algebra.
Hence, the Whittaker functions are in fact the eigenfunctions for the
open Toda chain. In the usual group-theoretical way the Whittaker function
is defined as a matrix element between compact and Whittaker vectors
\cite{Jac}-\cite{Ha} in the principal series representation.
There are many obstacles to generalize this approach to other quantum
models or to loop groups.

The present approach to construct the eigenfunctions for both periodic
and open chains is rather different \cite{KL1, KL2}: it bases on the
Quantum Inverse Scattering Method for the {\it periodic} Toda chain
\cite{Skl1}. One of the interesting results of analytical calculations
in the $R$-matrix framework is the revealing of recurrent relation
between $N$ and $N-1$ particles eigenfunctions for the {\it open} Toda
chain (in fact the idea to use a recurrent relation was pointed out by
Sklyanin in \cite{Skl3}; our recurrent relations is an explicit
realization of such an idea). This naturally leads
to new integral representation for the Weyl invariant Whittaker
functions to compare with classical results
\cite{Jac}-\cite{Ha}. This representation is quite explicit and very
useful to investigate the different asymptotics. In particular, the
Gindikin-Karpelevich formula \cite{GK} for the Harish-Chandra function
\cite{CH}  can be obtained in a very simple way for particular case
of $GL(N,\RR)$ group. The eigenfunction for the periodic
Toda chain are constructed in a rather explicit form and have
essentially the same form as the recurrent relation mentioned above.
The integral formula for eigenfunctions can be considered
as a representation of the Whittaker functions for $\wh{GL}(N)$ group
at the critical level.

The present approach can be generalized to other quantum integrable models.
For example, the eigenfunctions for the relativistic Toda chain are
calculated in \cite{KLS} using the same QISM ideology.

\section{Quantum Toda chain: description of the model}
\subsection{Periodic spectral problem}
The quantum $N$-periodic Toda chain is a multi-dimensional eigenvalue
problem with $N$ mutually commuting Hamiltonians
$H_k(x_1,p_1;\ldots;x_{\N},p_{\N})\,,\;(k=1,\ldots, N),$ where the
simplest Hamiltonians have the form
\be\label{hp0}
H_1=\sum_{k=1}^Np_k\\
H_2=\sum_{k<m}p_kp_m-\sum_{k=1}^Ne^{x_k-x_{k+1}}\\
H_3=\sum_{k<m<n}p_kp_mp_n+...,
\ee
$(x_{{\N}+1}\equiv x_1)$ etc. and the phase variables $x_k,p_k$ satisfy
the standard commutation relations $[x_k,p_m]=i\hbar\delta_{km}$.
The main goal is to find the solution to the eigenvalue problem
\be\label{sp0}
H_k\Psi_{\raise-3pt\hbox{$\scriptstyle\!\!\bfit E$}}\,=\,
E_k\Psi_{\raise-3pt\hbox{$\scriptstyle\!\!\bfit E$}}\hspace{1.5cm}
k=1,\ldots, N
\ee
with fast decreasing wave function
$\Psi_{\raise-3pt\hbox{$\scriptstyle\!\!\!\bfit E $}}$. To be more
precise, let us note that, due to translation invariance,
the solution to (\ref{sp0}) has the following structure:
\be\label{kv0}
\Psi_{\raise-3pt\hbox{$\scriptstyle\!\!\bfit E $}}(x_1,\ldots,x_{\N})=
\wt\Psi_{\raise-3pt\hbox{$\scriptstyle\!\!\bfit E $}}
(x_1\!-\!x_2,\ldots,x_{{\N}-1}\!-\!x_{\N})\,
\exp\left\{\frac{i}{\hbar}E_1\sum_{k=1}^Nx_k\right\}
\ee
One needs to find the solution to (\ref{sp3}) such that
$\wt\Psi_{\raise-2pt\hbox{$\scriptstyle\!\!\!\bfit E $}}
\in L^2(\RR^{N-1})$.
In equivalent terms, we impose the requirement
\be\label{kv}
\int f(E_1)\Psi_{\raise-3pt\hbox{$\scriptstyle\!\!\bfit E $}}
(x_1,\ldots,x_{N})dE_1\in L^2(\RR^N)
\ee
for any smooth function $f(y)\,,\;(y\in\RR)$ with finite support.

\subsection{$GL(N-1,\RR)$ spectral problem}
It turns out that solution to (\ref{sp0}), (\ref{kv}) can be
effectively written in terms of the wave functions corresponding to
{\it open} $N-1$-particle Toda chain (quantum $GL(N\!-\!1,\RR)$ chain).
The Hamiltonians of the latter system can be (formally) derived from
(\ref{hp0}) by cancelling out all the operators containing
$p_{\N}$ and $x_{\N}$ thus obtaining exactly
$N-1$ commuting Hamiltonians
$h_k(x_1,p_1;\ldots;x_{{\N}-1},p_{{\N}-1})$ ($k=1,\!\ldots,N-1$). Let
$\bgamma=(\gamma_1,\ldots,\gamma_{{\N}-1})\in\RR^{N-1}$,
$\bfit x=(x_1,\ldots,x_{{\N}-1})\in\RR^{N-1}$. We consider
$GL(N-1,\RR)$ spectral problem
\be\label{sp1}
\hspace{2cm}
h_k\psi_\bgamma(\bfit x)=\sigma_k(\bgamma)\psi_\bgamma(\bfit x)
\hspace{2cm}k=1,\ldots, N-1
\ee
where $\sigma_k(\bgamma)$ are elementary symmetric functions.

\bigskip\noindent
Obviously, in the asymptotic region
$x_{k+1}\gg x_k,\;(k=1,\ldots, N\!-\!2)$ all potentials vanish and the
solution to (\ref{sp1}) is a superposition
of plane waves. The problem is to find a solution to (\ref{sp1})
satisfying the following properties:
\begin{itemize}
\item[(i)] The solution vanishes very rapidly
\be\label{rpd}
\hspace{1cm}
\psi_\bgamma(\bfit x)\sim\exp\Big\{
-{\textstyle\frac{2}{\hbar}}\,e^{(x_k-x_{k+1})/2}\Big\}
\hspace{2cm}x_k-x_{k+1}\to\infty
\ee
\item[(ii)] The function $\psi_\bgamma$ is Weyl-invariant, i.e. it is
symmetric under any permutation
\be\label{we}
\psi_{...\gamma_j...\gamma_k...}=\psi_{...\gamma_k...\gamma_j...}
\ee
\item[(iii)]
$\psi_\bgamma$ can be analytically continued to an entire function of
$\bgamma\in\CC^{N-1}$ and the following asymptotics hold:  \be\label{as2}
\psi_\bgamma\,\sim\,|\gamma_j|^{\frac{2-N}{2}}
\exp\Big\{-\frac{\pi}{2\hbar}(N\!-\!2)|\gamma_j|\Big\}
\ee
as $|{\rm Re}\,\gamma_j|\to\infty$ in the finite strip of complex
plane.
\end{itemize}
The properties (i)-(iii) define a unique solution to the spectral
problem (\ref{sp0}).

\section{Main results}
\begin{th}\label{th0} The following statements hold \cite{KL1, KL2}:
\begin{itemize}
\item[(i)]
Let a set $||\gamma_{jk}||$ be the lower triangular
$(N\!-\!1)\times(N\!-\!1)$ matrix.
The solution to the spectral problem
(\ref{sp1})-(\ref{as2}) can be written
in the form of multiple
Mellin-Barnes integrals
\footnote{We identify the set
$\bgamma$ with the last row
$(\gamma_{{\N}-1,1},\ldots,\gamma_{{\N}-1,{\N}-1})$.
}:
\be\label{mb}
\psi_{\gamma_{_{\scriptstyle\N-1,1}},
\ldots,\gamma_{_{\scriptstyle\N-1,\N-1}}}
(x_1,\ldots ,x_{{\N}-1})\;=\\ =
\frac{(2\pi\hbar)^{-\frac{(N-1)(N-2)}{2}}}{\prod_{k=1}^{N-2}k!}
\int\limits_{\cal C} \prod_{n=1}^{N-2}
\frac{\prod_{j=1}^n\prod_{k=1}^{n+1}
\hbar^{\frac{\scriptstyle\gamma_{nj}-\gamma_{n+1,k}\!}
{\scriptstyle i\hbar}}\,
\Gamma\Big(\frac{\textstyle\gamma_{nj}
\!-\!\gamma_{n+1,k}}{\textstyle i\hbar}\Big)}
{\prod\limits_{\stackrel{\scriptstyle j,k=1}{j<k}}^n
\Gamma\Big(\frac{\textstyle\gamma_{nj}-\gamma_{nk}}
{\textstyle i\hbar}\Big)
\Gamma\Big(\frac{\textstyle\gamma_{nk}-\gamma_{nj}}
{\textstyle i\hbar}\Big)}\;\times \\
\exp\left\{\frac{i}{\hbar}\sum_{n,k=1}^{N-1}x_n
\Big(\gamma_{nk}-\gamma_{n-1,k}\Big)\right\}
\prod\limits_{\stackrel{\scriptstyle j,k=1}{j\leq k}}^{N-2}
d\gamma_{jk}
\ee
where the integral should be understand as follows:
first we integrate on  $\gamma_{11}$ over the line
${\rm Im}\,\gamma_{11} >
\max\{{\rm Im}\,\gamma_{21}, {\rm Im}\,\gamma_{22}\}$;
then we integrate on the set $(\gamma_{21} ,\gamma_{22})$ over
the lines ${\rm Im}\,\gamma_{2j} >
\max_{m}\{{\rm\,Im}\gamma_ {3m}\}$ and so on. The last integrations
should be performed on the set of variables
$(\gamma_{{\N}-2,1}\ldots,\gamma_{{\N}-2,{\N}-2})$
over the lines
${\rm Im}\gamma_{{\N}-2,k}>\max_{m}\{{\rm Im}\,\gamma_{{\N}-1,m}\}$.
\item[(ii)]
In the region $x_k\ll x_{k+1} (k=1,\ldots, N-1)$ the solution has
the following asymptotics:
\be\label{ch}
\psi_\bgamma(\bfit x)=
\sum_{s\in W}\phi(s\bgamma)
e^{\frac{i}{\hbar}(s\bgamma,\bfit x)}+
O\Big(\mbox{\rm max}\Big\{e^{x_k-x_{k+1}}\Big\}_{k=1}^{\N-1}\Big)
\ee
where $(.,.)$ is a scalar product in $\RR^{N-1}$ and
the summation is performed over the permutation group;
$\phi(\bgamma)$ is (renormalized) Harish-Chandra function
\be\label{ch2}
\phi(\bgamma)=\hbar^{-2i(\bgamma,\brho)/\hbar}
\prod_{j<k}\Gamma\Big(\frac{\gamma_j\!-\!\gamma_k}{i\hbar}\Big)
\ee
where $(\bgamma,\brho)\equiv
\frac{1}{2}\sum\limits_{m=1}^{N-1}(N-2m)\gamma_k$.
\item[(iii)]
The functions (\ref{mb}) have the scalar product
\be\label{ort}
\hspace{1cm}
\int\limits_{\rs^{N-1}}\ov\psi_{\bgamma'}(\bfit x)\psi_\bgamma(\bfit x)
d\bfit x=\frac{\mu^{-1}(\bgamma)}{(N\!-\!1)!}\,\sum_{s\in W}
\delta(s\bgamma-\bgamma')
\hspace{1cm}(\bgamma,\bgamma'\in\RR^{N-1})
\ee
and obey the completeness condition
\be\label{com}
\int\limits_{\rs^{N-1}}\mu(\bgamma)\psi_\bgamma(\bfit x)
\ov\psi_\bgamma(\bfit y)d\bfit\bgamma=\delta(\bfit x-\bfit y)
\ee
where
\be\label{meas}
\mu(\bgamma)=\frac{(2\pi\hbar)^{1-N}}{(N\!-\!1)!}\,\prod_{j<k}
\left|\Gamma\Big(\frac{\gamma_j\!-\!\gamma_k}{i\hbar}\Big)\right|^{-2}
\ee
is the Sklyanin measure \cite{Skl1}.
\end{itemize}
\end{th}
The eigenfunctions for the periodic chain are constructed as a kind
of Fourier transform with the function (\ref{mb}). Let
\be\label{gsz}
t_{\N}(\la;\bfit E)=\sum\limits_{k=0}^N(-1)^k\la^{N-k}E_k
\ee
and $\bfit e_j$ denotes $j$-th basis vector in $\RR^{N-1}$.
\begin{th}\label{th1}
The solution to the spectral problem (\ref{sp0}), (\ref{kv}) can be
represented as the integral over real variables
$\bgamma=(\gamma_1,\ldots,\gamma_{\N-1})$ in the following form:
\be\label{m1}
\Psi_{\raise-3pt\hbox{$\scriptstyle\!\!\bfit E $}}(\bfit x,x_{\N})=
\frac{1}{2\pi}\;\int\limits_{\rs^{N-1}}\mu
(\bgamma)C(\bgamma;\bfit E)\,\Psi_{\bgamma,E_1}(\bfit x,x_{\N})d\bgamma
\ee
where
\begin{itemize}
\item[(i)] The function $\Psi_{\bgamma, E_1}(\bfit x,x_{\N})$
is defined in terms of solution (\ref{mb}) to the $GL(N\!-\!1,\RR)$
spectral problem:
\be\label{m6}
\Psi_{\bgamma, E_1}(\bfit x,x_{\N})=
\,\psi_\bgamma(\bfit x)\,
\exp\Big\{\frac{i}{\hbar}\Big(E_1\!-\!\sum_{m=1}^{N-1}\gamma_m\Big)
x_N\Big\}
\ee
\item[(ii)] The function $C(\bgamma;\bfit E)$ is the
solution of multi-dimensional Baxter equations
\be\label{b1}
t_{\N}(\gamma_j;\bfit E)C(\bgamma;\bfit E)=
i^NC(\bgamma+i\hbar\bfit e_j;\bfit E)
+i^{-N}C(\bgamma-i\hbar\bfit e_j;\bfit E)
\ee
which is symmetric entire function in
$\bgamma$-variables with the asymptotics
\be\label{as1}
C(\bgamma;\bfit E)\sim
|\gamma_k|^{-N/2}\exp\Big\{-\frac{\pi N|\gamma_k|}{2\hbar}\Big\}
\ee
as ${\rm Re}\,\gamma_k\to\pm\infty$ in the strip
$|{\rm Im}\,\gamma_k|\leq\hbar$
\end{itemize}
\end{th}

\bigskip\noindent
The above restrictions imposed on solution to (\ref{b1}) are
reformulation of the quantization condition (\ref{kv}) on the level
of $\bgamma$-representation. To obtain the explicit integral form for
the eigenfunctions, we use the solution to (\ref{b1}), (\ref{as1})
in the Pasquier-Gaudin form \cite{PG} (see sect.7 below)
\be\label{pg1}
C(\bgamma;\bfit E)=\prod_{j=1}^{N-1}\frac{c_+(\gamma_j;\bfit E)-
\xi(\bfit E)c_-(\gamma_j;\bfit E)}
{\prod\limits_{k=1}^N\sinh\frac{\textstyle\pi}{\textstyle\hbar}
\Big(\gamma_j-\delta_k(\bfit E)\Big)}
\ee
where the entire functions $c_\pm(\gamma)$ are two Gutzwiller's
solutions \cite{Gu} of the one-dimen\-sio\-nal Baxter equation
\be
t(\gamma;\bfit E)c(\gamma;\bfit E)=i^{-N}c(\gamma+i\hbar;\bfit E)+
i^{N}c(\gamma-i\hbar;\bfit E) \ee and the parameters $\xi(\bfit E),\;
\bfit\delta=(\delta_1(\bfit E),\ldots, \delta_{\N}(\bfit E)\,)$
satisfy the Gutzwiller conditions (the energy quantization)
\cite{Gu, PG}(see sect.7 below). Then the multiple integral (\ref{m1})
can be explicitly evaluated. Let $\bfit y=(y_1,\ldots,y_{\N})\!\in\RR^N$
be an arbitrary vector. We denote
$\bfit y^{(s)}\equiv(y_1,\ldots,y_{s-1},y_{s+1},\ldots,y_{\N})$
the corresponding vector in $\RR^{N-1}$.
\begin{th}\label{th2}
Assuming that $\delta_j(\bfit E)\neq \delta_k(\bfit E)$, the solution
(\ref{m1}) can be written (up to an inessential numerical factor) in the
equivalent form
\be\label{gu0}
\Psi_{\raise-3pt\hbox{$\scriptstyle\!\!\bfit E $}}(\bfit x,x_{\N})=
\sum_{s=1}^N(-1)^{N-s}\!\!\sum_{\bfit n^{(s)}\in\zs^{\N-1}}
\!\!\Delta(\bfit\delta^{(s)}\!+\!i\hbar\bfit n^{(s)})\,
C_+(\bfit\delta^{(s)}\!+\!i\hbar\bfit n^{(s)})\,
\Psi_{\bfit\delta^{(s)}+i\hbar\bfit n^{(s)},E_1}(\bfit x,x_{\N})
\ee
where
\be
C_+(\bgamma)\equiv\prod_{j=1}^{N-1}c_+(\gamma_j;\bfit E)
\ee
and $\Delta(\bgamma)=\prod\limits_{j>k}(\gamma_j-\gamma_k)$ is the
Vandermonde determinant.
\end{th}
\begin{rem}
For $N=2, 3$ and $N=4$  formula (\ref{gu0})
reproduces
the results obtained by Gutzwiller \cite{Gu}.
\end{rem}

\section{$R$-matrix approach}
The Toda chain can be nicely described using the $R$-matrix approach
\cite{Skl1}.
It is well known that the Lax operator
\be\label{ln}
L_n(\la)\,=\, \left(\begin{array}{cc}\la-p_n & e^{-x_n}\\ -e^{x_n} & 0
\end{array}\right)
\ee
satisfies the commutation relations
\be
R(\la-\mu)(L_n(\la))\otimes I)(I\otimes L_n(\mu))=
(I\otimes L_n(\mu))(L_n(\la)\otimes I)R(\la-\mu)
\ee
where
\be
R(\la)=I\otimes I+\frac{i\hbar}{\la}\,P
\ee
is a rational $R$-matrix. The monodromy matrix
\be\label{a1}
T_{_N}(\la)\;\stackrel{\mbox{\tiny def}}{=}\;L_{\N}(\la)\ldots
L_1(\la)\,\equiv\, \left(\begin{array}{cc}A_{\N}(\la) & B_{\N}(\la)\\
C_{\N}(\la) & D_{\N}(\la)\end{array}\right)
\ee
satisfies the analogous equation
\be\label{rtt}
R(\la-\mu)(T(\la)\otimes I)(I\otimes T(\mu))=
(I\otimes T(\mu))(T(\la)\otimes I)R(\la-\mu)
\ee
In particular, the following commutation relations hold:
\be\label{m2}
[A_{\N}(\la),A_{\N}(\mu)]=[C_{\N}(\la),C_{\N}(\mu)]=0
\ee
\be\label{m3}
(\la-\mu+i\hbar)A_{\N}(\mu)C_{\N}(\la)\,=\,
(\la-\mu)C_{\N}(\la)A_{\N}(\mu)+i\hbar A_{\N}(\la)C_{\N}(\mu)
\ee
\be\label{m4}
(\la-\mu+i\hbar)D_{\N}(\la)C_{\N}(\mu)\,=\,
(\la-\mu)C_{\N}(\mu)D_{\N}(\la)+i\hbar D_{\N}(\mu)C_{\N}(\la)
\ee
From (\ref{rtt}) it can be easily shown that the trace of the monodromy
matrix
\be\label{tra}
\wh t_{\N}(\la)=A_{\N}(\la)+D_{\N}(\la)
\ee
satisfies the commutation relations $[\wh t(\la),\wh t(\mu)]=0$ and
is a generating function for the Hamiltonians of the periodic Toda chain:
\be
\wh t_{\N}(\la)=\sum_{k=0}^N(-1)^k\la^{N-k}H_k
\ee
We reformulate the spectral equations (\ref{sp0}) as follows:
\be\label{sp3}
\wh t_{\N}(\la)\Psi_{\raise-3pt\hbox{$\scriptstyle\!\!\bfit E $}}=
t_{\N}(\la;\bfit E)\Psi_{\raise-3pt\hbox{$\scriptstyle\!\!\bfit E $}}
\ee
where
\be\label{eig}
t_{\N}(\la;\bfit E)=\sum\limits_{k=0}^N(-1)^k\la^{N-k}E_k
\ee
On the other hand, it can be easily shown that the operator
\be\label{au1}
A_{{\N}-1}(\la)\,\equiv\,\sum_{k=0}^{N-1}(-1)^k\la^{N-k-1}
h_k(x_1,p_1;\ldots;x_{\N-1},p_{\N-1})
\ee
is nothing but the generating function for the Hamiltonians $h_k$ of
$GL(N\!-\!1)$ Toda chain. Therefore, the $GL(N-1,\RR)$ spectral
equations can be written in the form
\be\label{au4}
A_{{\N}-1}(\la)\psi_\bgamma(\bfit x)=\prod_{m=1}^{N-1}(\la-\gamma_m)\,
\psi_\bgamma(\bfit x)
\ee
Using the obvious relation
\be\label{ca}
C_{\N}(\la)=-e^{x_N}A_{{\N}-1}(\la)
\ee
one obtains, as a trivial corollary of (\ref{au4}),
\be\label{au5}
\hspace{2cm}
C_{\N}(\gamma_j)\psi_\bgamma(\bfit x)=0
\hspace{2cm}\forall\,\gamma_j\in\bgamma
\ee
\begin{rem}
Equations (\ref{au5}) are an analytical analog of the notion of
"operator zeros" introduced by Sklyanin \cite{Skl1}.
\end{rem}

\section{Eigenfunctions for the open Toda chain}
Suppose that solution to (\ref{au4}) satisfying (\ref{rpd})-(\ref{as2})
is given. Using the commutation relations (\ref{m3}), (\ref{m4})
together with (\ref{au5}), it is easy to show that the following
relations hold
\bse\label{o1}
\be\label{o1a}
A_{\N}(\gamma_j)\psi_\bgamma=i^{-N}e^{-x_N}
\psi_{\bgamma-i\hbar\bfit e_j}
\ee
\be\label{o1b}
D_{\N}(\gamma_j)\psi_\bgamma=i^Ne^{x_N}
\psi_{\bgamma+i\hbar\bfit e_j}
\ee
\ese
($j=1,\ldots,N-1$) where $\bfit e_j$ is $j$-th basis vector
in $\RR^{N-1}$. Note that (\ref{o1b}) is a corollary of
(\ref{o1a}) since the quantum determinant of the monodromy
matrix (\ref{a1}) is unity.

\bigskip\noindent
Let us introduce the key object - the {\it auxiliary} function
\be\label{m10}
\Psi_{\bfit\gamma,\e}(x_1,\ldots, x_{\N})
\;\stackrel{\mbox{\tiny def}}{=} \;\psi_{\bfit\gamma} (\bfit x)
\exp\Big\{\frac{i}{\hbar}
\Big(\e\,-\!\sum\limits_{m=1}^{N-1}\gamma_m\Big)x_{\N}\Big\}
\ee
where $\e$ is an arbitrary parameter. From (\ref{au4}), (\ref{ca})
and (\ref{o1}) it is readily seen that this
function satisfies to equations
\bse\label{int}
\be\label{m13}
C_{\N}(\la)\Psi_{\bgamma,\e}=-\,e^{x_N}\prod_{j=1}^{N-1}
(\la-\gamma_j)\;\Psi_{\bgamma,\e}
\ee
\be\label{m11}
A_{\N}(\la)\Psi_{\bgamma,\e}=
\Big(\la-\e+\sum_{m=1}^{N-1}\gamma_m\Big)
\prod_{j=1}^{N-1}(\la\!-\!\gamma_j)\;\Psi_{\bgamma,\e}+
i^{-N}\sum_{j=1}^{N-1}\Psi_{\bfit\gamma-i\hbar\bfit e_j,\e}\,
\prod_{m\neq j}\frac{\la\!-\!\gamma_m}{\gamma_j\!-\!\gamma_m}
\ee
\be\label{m12}
D_{\N}(\la)\Psi_{\bgamma,\e}=
i^N\sum_{j=1}^{N-1}\Psi_{\bfit\gamma+i\hbar\bfit e_j,\e}\,
\prod_{m\neq j}\frac{\la\!-\!\gamma_m}{\gamma_j\!-\!\gamma_m}
\ee
\ese
In particular,
\be\label{m14}
\wh t_{\N}(\gamma_j)\Psi_{\bgamma,\e}=i^{N}\Psi_{\bgamma,\e}+
i^{-N}\Psi_{\bgamma,\e}
\ee

\bigskip\noindent
The problem is to find the corresponding solution
for $GL(N,\RR)$ Toda chain using the above information, i.e.
in terms of the function $\Psi_{\bgamma,\e}(\bfit x)$
construct the Weyl invariant function
$
\psi_{\la_1,\ldots,\la_{\N}}(x_1,\ldots,x_{\N})
$
satisfying to equations
\bse\label{p1}
\be\label{p1a}
A_{\N}(\la)\psi_{\la_1,\ldots,\la_N}=
\prod_{k=1}^N(\la-\la_k)\;\psi_{\la_1,\ldots,\la_N}
\ee
\be\label{p1b}
\hspace{1.5cm}
A_{{\N}+1}(\la_n)\psi_{\la_1,\ldots,\la_N}=
i^{-N-1}\,e^{x_{N+1}}\psi_{\la_1,\ldots,\la_n-i\hbar,\ldots,\la_N}
\hspace{1cm}(n=1,\ldots, N)
\ee
\ese
and obeying the similar to (\ref{rpd})-(\ref{as2}) conditions
\begin{lem}\label{lem2}\cite{KL2}
Let $\Psi_{\bgamma,\e}(\bfit x,x_{\N})$ be the auxiliary function
(\ref{m10}). Let $\blambda=(\la_1,\ldots,\la_{{\N}})\!\in\CC^N\!$
be the set of indeterminates. Let
\be\label{re2'}
\mu(\bgamma)=\frac{(2\pi\hbar)^{1-N}}{(N\!-\!1)!}\prod_{j<k}
\left\{\Gamma\Big(\frac{\gamma_j\!-\!\gamma_k}{i\hbar}\Big)
\Gamma\Big(\frac{\gamma_k\!-\!\gamma_j}{i\hbar}\Big)\right\}^{-1}
\ee
\be\label{re2}
Q(\gamma_1,\ldots,\gamma_{{\N}-1}|\lambda_1,\ldots,\lambda_{\N})=
\prod_{j=1}^{N-1}\prod_{k=1}^N
h^{\frac{\scriptstyle\gamma_j-\la_k}{\scriptstyle i\hbar}}\,
\Gamma\Big(\frac{\gamma_j\!-\!\la_k}{i\hbar}\Big)
\ee
Then the Weyl invariant solution to the spectral problem
(\ref{p1a})-(\ref{p1b}) with the properties discussed above is given by
recurrent formula
\be\label{re4}
\psi_{\la_1,\ldots,\la_N}(x_1,\ldots,x_{\N})=
\int\limits_{\cal C}
\mu(\bgamma)Q(\bgamma;\blambda)\Psi_{\bfit\gamma;
\la_1+\ldots+\la_N}(x_1,\ldots,x_{\N})\,d\bgamma
\ee
where the integration is performed along the horizontal lines with
$\mbox{\rm Im}\,\gamma_j>\mbox{\rm max}_k\,\{\mbox{\rm Im}\,\la_k\}$.
\end{lem}
{\bf Proof.} One needs to calculate the action of the operators
$A_{\N}(\la)$ and
\be
A_{{\N}+1}(\la)=(\la-p_{\N+1})A_{\N}(\la)+e^{-x_{N+1}}C_{\N}(\la)
\ee
on the function (\ref{re4}) using the formulae (\ref{m11}) and
(\ref{p1a}), (\ref{m13}). The shifted contours can be deformed to
original ones using the facts that integrand in (\ref{re4}) is an
entire function fast decreasing in any finite horizontal strip of
complex plane as $|{\rm Re}\,\gamma_j|\to\infty$. The last step is to
use the difference equations for the parts of the integrand
with respect to the shifts $\pm i\hbar$ of parameters $\gamma_m$ and
$\lambda_k$. \square

\bigskip\noindent
{\bf Proof of Theorem \ref{th0}.}
The proof of (\ref{mb}) is straightforward resolution of the
recurrent relations (\ref{re4}) starting with trivial eigenfunction
$\psi_{\gamma_{_{11}}}(x_1)=\exp\{\frac{i}{\hbar}\gamma_{11} x_1\}$.
Obviously, the function (\ref{mb}) is symmetric under the permutation of
parameters $\bgamma$. The asymptotics (\ref{rpd}) can be proved using
the steepest descent method. Using the Stirling formula for the
$\Gamma$-functions as $\gamma_{{\N}-1,k}\equiv\gamma_k\to\pm\infty$,
it is easy to see that the asymptotics (\ref{as2}) hold. Hence,
(\ref{mb}) is an appropriate solution to the spectral problem.\\
Further, the formula (\ref{ch}) can be proved as follows.
The integrand in (\ref{re4}) decreases exponentially as
$|\gamma_j|\to\infty\,,\,(j=1,\ldots,N\!-\!1)$ in the lower half-plane
and, as consequence, the integrals over large semi-circles in the
lower half-plane vanish. Using the Cauchy formula to calculate
the integral (\ref{re4}) in the asymptotic region
$x_{k+1}\gg x_k,\;(k=1,\ldots,N\!-\!1)$,
it is easy to see that the asymptotics of the function
$\psi_{\bgamma}$ are determined precisely in terms of
the corresponding Harish-Chandra function (\ref{ch2}).\\
The scalar product (\ref{ort}) is the consequence of
the Plancherel formula proved in \cite{ST} for the $SL(N,\RR)$ case.
The formula (\ref{com}) can be proved by induction. \square
\begin{rem}
In \cite{KL1} (eqs.(4.7),(4.18)) the eigenfunctions for an open
Toda chain have been constructed in terms of Whittaker
function which has a standard integral representation corresponding
to the Iwasawa decomposition of semi-simple group
(see for example \cite{Ha}). Our expression (\ref{mb}), being obtained
in the framework of Quantum Inverse Scattering Method, seems
quite different. Nevertheless, both representations do define the same
function (this can be shown by comparing the corresponding asymptotics
and analytical properties). Hence, one can consider the representation
(\ref{mb}) as a new one for Whittaker function \cite{Jac}-\cite{Ha}.
\end{rem}

\section{Periodic chain: $\bgamma$-representation,
eigenfunctions, \\ and Plancherel formula}
Let $\Psi_{\raise-2pt\hbox{$\scriptstyle\!\!\!\bfit E $}}(\bfit x,x_{\N})$
be the fast decreasing solution of the problem (\ref{sp3}).
We define the function $C(\bgamma;\bfit E)$ by the generalized Fourier
transform:
\be\label{c1}
\delta(E_1-\e)\,C(\bgamma;\bfit E)=\int\limits_{\rs^{N-1}}
\Psi_{\raise-3pt\hbox{$\scriptstyle\!\!\bfit E $}}(\bfit x,x_{\N})
\,\ov\Psi_{\bfit\gamma,\e}(x_0,\bfit x)d\bfit xdx_{\N}
\ee
\begin{lem}\label{lem1}
The function $C(\bgamma)$ possesses the following properties:
\begin{itemize}
\item[(i)] It is a symmetric function with respect to $\bgamma$-variables.
\item[(ii)] It is an entire function of  $\bgamma\in\CC^{N-1}$.
\item[(iii)] The function $C(\bgamma)$ obeys the asymptotics
\be\label{as4}
C(\bgamma;\bfit E)\sim
|\gamma_k|^{-N/2}\exp\Big\{-\frac{\pi N|\gamma_k|}{2\hbar}\Big\}
\ee
as ${\rm Re}\,\gamma_k\to\pm\infty$ in the strip
$|{\rm Im}\,\gamma_k|\leq\hbar$.
\item[(iv)] The function $C(\bgamma)$ satisfies the multi-dimensional
Baxter equation
\be\label{b11}
t(\gamma_j;\bfit E)C(\bgamma;\bfit E)=
i^NC(\bgamma+i\hbar\bfit e_j;\bfit E)
+i^{-N}C(\bgamma-i\hbar\bfit e_j;\bfit E)
\ee
where $t(\gamma;\bfit E)$ is defined by (\ref{eig}).
\end{itemize}
\end{lem}
{\bf Proof.}
The symmetry of the function $C(\bgamma)$ is obvious.\\
We present here only a sketch of the proof of the statements
(ii) and (iii).\\
The statement (ii) follows from the assertion that the auxiliary function
$\Psi_{\bgamma,\e}$ is an entire one while the solution of the
periodic chain vanishes very rapidly as $|x_k\!-\!x_{k+1}|\to\infty$.
\footnote{
Actually, the boundary conditions have the same importance here as the
requirement of compact support in the theory of the analytic
continuation for the usual Fourier transform.
}.\\
(iii) The asymptotics (\ref{as4}) is a combination of two factors.
The first one comes from the asymptotics (\ref{as2}) while the
additional factor
$\sim |\gamma_k|^{-1}\exp\{-\pi|\gamma_k|/\hbar\}$ results
from the stationary phase method while calculating the multiple integral
including the function (\ref{mb}). The calculation is based
heavily upon the exact asymptotics of the function
$\Psi_{\raise-2pt\hbox{$\scriptstyle\!\!\!\bfit E $}}(\bfit x,x_{\N})$ as
$|x_k-x_{k+1}|\to\infty$.\\
The proof of (iv) is simple. Using the definition (\ref{sp3}) and
integrating by parts (evidently, boundary terms vanish), one obtains
\be
\delta(E_1-\e)\,t(\gamma_j;\bfit E)C(\bgamma)\equiv
\int\limits_{\rs^{N-1}}
\Big\{\wh t(\gamma_j)\Psi_{\raise-3pt\hbox{$\scriptstyle\!\!\bfit E
$}}(\bfit x,x_{\N})\Big\}
\,\ov\Psi_{\bfit\gamma,\e}(\bfit x,x_{\N})d\bfit xdx_{\N} =\\
=\int\limits_{\rs^{N-1}}
\Psi_{\raise-3pt\hbox{$\scriptstyle\!\!\bfit E $}}(\bfit x,x_{\N})
\,\ov{\wh t(\gamma_j)\Psi_{\bfit\gamma,\e}(\bfit x,x_{\N})}
d\bfit xdx_{\N}
\ee
Taking into account the relation (\ref{m14}), the Baxter equation
(\ref{b11}) follows from definition (\ref{c1}).\square

\bigskip\noindent
Now we prove Theorem \ref{th1}. Using the completeness condition
\be\label{con2}
\int\limits_{\rs^{N}}\mu(\bgamma)\Psi_{\bgamma,\e}(\bfit x,x_{\N})
\ov\Psi_{\bgamma,\e}(\bfit y,y_{\N})d\bgamma d\e\,=\,
2\pi\hbar\delta(\bfit x-\bfit y)\delta(x_{\N}-y_{\N})
\ee
which is a corollary of (\ref{com}), the inversion of the formula
(\ref{c1}) results to expression
\be\label{m1'}
\Psi_{\raise-3pt\hbox{$\scriptstyle\!\!\bfit E $}}(\bfit x,x_{\N})=
\frac{1}{2\pi}\;\int\limits_{\rs^{N-1}}\mu
(\bgamma)C(\bgamma;\bfit E)\,\Psi_{\bgamma,E_1}(\bfit x,x_{\N})d\bgamma
\ee
The integral (\ref{m1'}) is correctly defined. Indeed, the
measure (\ref{meas}) is an entire function. Therefore, there are no
poles in the integrand. Moreover,
\be\label{mu2}
\mu(\bgamma)\sim|\gamma_k|^{N-2}
\exp\Big\{\frac{\pi}{\hbar}(N\!-\!2)|\gamma_k|\Big\}
\ee
as $|\gamma_k|\to\infty$. Taking into account the asymptotics (\ref{as2})
and (\ref{as4}) one concludes that the integrand has the behavior
$\sim|\gamma_k|^{-1}\exp\{-\pi|\gamma_k|/\hbar\}$ as
$|\gamma_k|\to\infty$.
Therefore, the integral (\ref{con2}) is convergent.\\
One can directly prove the spectral problem (\ref{sp3}) calculating
the action of the operator $\wh t_{\N}(\la)=A_{\N}(\la)+D_{\N}(\la)$
on the right hand side of (\ref{m1'}) with the help of the formulae
(\ref{m11}), (\ref{m12}). The calculation is performed similarly to
those of Lemma \ref{lem2}, using the analytical properties of the
integrand and the Baxter equation (\ref{b11}) (see \cite{KL1} for
details).

\bigskip\noindent
The last step is to prove that the function (\ref{m1'})
satisfies to integrability requirement (\ref{kv}).
Using the scalar product
\be\label{pl2}
\int\limits_{\rs^{N}}\ov\Psi_{\bfit\gamma',\e'}(\bfit x,x_{\N})
\Psi_{\bfit\gamma,\e}(\bfit x,x_{\N})d\bfit xdx_{\N}=
(2\pi\hbar)\frac{\mu(\bgamma)}{(N\!-\!1)!}\,
\delta(\e\!-\!\e')\sum_{s\in W}\delta(s\bgamma-\bgamma')
\ee
one can write the Plancherel formula
\be\label{top}
2\pi\hbar\int\limits_{\rs^{N}}
\ov\Psi_{\raise-3pt\hbox{$\scriptstyle\!\!\bfit E'$}}(\bfit x,x_{\N})
\Psi_{\raise-3pt\hbox{$\scriptstyle\!\!\bfit E $}}(\bfit x,x_{\N})
d\bfit xdx_{\N} =\delta(E_1\!-\!E_1')\!
\int\limits_{\rs^{N-1}}\!\!\mu(\bgamma)\,\ov C(\bgamma;\bfit E')
C(\bgamma;\bfit E)d\bgamma
\ee
The integral in the r.h.s. of (\ref{top}) is absolutely convergent due to
asymptotics (\ref{as4}) and (\ref{mu2}). Hence, the norm
$||\Psi_{\raise-2pt\hbox{$\scriptstyle\!\!\!\bfit E $}}||$
is finite modulo $GL(1)$ $\delta$-function $\delta(E_1\!-\!E_1')$
(see the corresponding factor in (\ref{kv0}) which leads to this
function) and requirement (\ref{kv}) is fulfilled.
Hence, Theorem \ref{th1} is proved. \square

\section{Solution of Baxter equation}
It is well known \cite{Gu, PG} (see also \cite{KL1} for details)
that the solution to the Baxter
equation (\ref{b1}) with the asymptotics (\ref{as1}) can be written
in the following separated form:
\be\label{be1}
C(\bgamma;\bfit E)=\prod_{j=1}^{N-1}\frac{c_+(\gamma_j;\bfit E)-
\xi c_-(\gamma_j;\bfit E)}
{\prod\limits_{k=1}^N\sinh\frac{\textstyle\pi}{\textstyle\hbar}
\big(\gamma_j-\delta_k)}
\ee
where $\xi$ and $\delta_k$ are an arbitrary constants and the entire
functions $c_\pm(\gamma)$ are defined it terms of $\NN\times\NN$
determinants:
\bse\label{b2}
\be\label{b2a}
c_+(\gamma)=\hbox{\footnotesize $
\frac{\displaystyle1}{\textstyle\prod\limits_{k=1}^N
\hbar^{\!\raise2pt\hbox{$ {\scriptstyle-}\frac{i\gamma}{\hbar} $}}
\Gamma(1\!-\!\frac{i}{\hbar}(\gamma\!-\!\la_k))}
\left|\!\!
\begin{array}{cccccc}
 1 &\frac{1}{t(\gamma\!+\!i\hbar)} & 0 & \ldots & \ldots &
...
\\
\frac{1}{t(\gamma\!+\!2i\hbar)} &1 & \frac{1}{t(\gamma\!+\!2i\hbar)}
& 0 & \ldots & \ldots
\\
0 &\frac{1}{t(\gamma\!+\!3i\hbar)} &1
&\frac{1}{t(\gamma\!+\!3i\hbar)}& 0  & \ldots
\\
\ldots &\ldots &\ldots & \ldots & \ldots & \ldots
\end{array}
\!\!\right|$}
\ee
\be\label{b2b}
c_-(\gamma)=\hbox{\footnotesize $
\frac{\displaystyle 1}{\textstyle\prod\limits_{k=1}^N
\hbar^{\!\raise2pt\hbox{${\scriptstyle}\frac{i\gamma}{\hbar}$}}
\Gamma(1+\frac{i}{\hbar}(\gamma-\la_k))}
\left|\!\!
\begin{array}{cccccc}
\ldots  & \ldots & \ldots & \ldots & \ldots & \ldots\\
\ldots & 0& \frac{1}{t(\gamma\!-\!3i\hbar)}& 1 &
\frac{1}{t(\gamma\!-\!3i\hbar)} & 0\\
\ldots &\ldots  & 0 &\frac{1}{t(\gamma\!-\!2i\hbar)} & 1  &
\frac{1}{t(\gamma\!-\!2i\hbar)}\\
\ldots &\ldots & \ldots & 0 & \frac{1}{t(\gamma\!-\!i\hbar)} & 1
\end{array}
\!\!\right|$}
\ee
\ese
and $\la_k\equiv\la_k(\bfit E)$ are the roots of the polynomial
$t(\gamma)\equiv t_{\N}(\gamma;\bfit E)$.\\
On the other hand, the solution (\ref{be1}) is not an
entire function in general since the denominator in (\ref{be1})
has an infinite number of poles at $\gamma=\delta_k+i\hbar
n_k\,,\;n_k\in\ZZ,\,k=1,\ldots, N$. The poles are cancelled only
if the following conditions hold:
\be\label{be3}
c_+(\delta_k+i\hbar n_k)=\xi c_-(\delta_k+i\hbar n_k)
\ee
In turn, this means that the Wronskian
\be\label{be4}
W(\gamma)=c_+(\gamma)c_-(\gamma+i\hbar)-
c_+(\gamma+i\hbar)c_-(\gamma)
\ee
vanishes at $\gamma=\delta_k+i\hbar n_k$. The Wronskian is
$i\hbar$-periodic function and possesses exactly $N$ real roots
$\delta_k(\bfit E)$ \cite{Gu}. Therefore, the solution (\ref{be1})
has no poles if one takes $\delta_k=\delta_k(\bfit E)$
provided that the constant $\xi$ is chosen in such a way that
\be\label{xi}
\hspace{2cm}
\xi=\left.\frac{c_+(\gamma)}{c_-(\gamma)}\right|_{\gamma=\delta_k(\bfit E)}
\hspace{2cm}k=1,\ldots,N
\ee
Hence, one arrives at the following
\begin{lem}\cite{PG}
The function
\be\label{pg2}
C(\bgamma;\bfit E)=\prod_{j=1}^{N-1}\frac{c_+(\gamma_j;\bfit E)-
\xi(\bfit E)c_-(\gamma_j;\bfit E)}
{\prod\limits_{k=1}^N\sinh\frac{\textstyle\pi}{\textstyle\hbar}
\Big(\gamma_j-\delta_k(\bfit E)\Big)}
\ee
where $\delta_k(\bfit E)$ are real zeros of the Wronskian
(\ref{be4}) and the constant $\xi$ is chosen according to (\ref{xi}),
satisfies to conditions of Lemma \ref{lem1}.
\end{lem}
The quantization conditions
\be\label{q2}
\frac{c_+(\delta_1)}{c_-(\delta_1)}=\ldots=
\frac{c_+(\delta_{\N})}{c_-(\delta_{\N})}
\ee
determine the energy spectrum of the problem. They have been obtained for
the first time by Gutzwiller \cite{Gu} using quite different method.

\bigskip\noindent
To prove Theorem \ref{th2}, one should substitute the solution (\ref{pg2})
into the integral formula (\ref{m1}) and calculate the residues
coming from individual terms
\be\label{it}
\frac{c_\pm(\gamma_j;\bfit E)}
{\prod\limits_{k=1}^N\sinh\frac{\textstyle\pi}{\textstyle\hbar}
\Big(\gamma_j-\delta_k(\bfit E)\Big)}
\ee
The result is exactly the sum over all possible poles of expressions
(\ref{it}) and essentially coincides with (\ref{gu0}) (see careful
analysis in \cite{KL1}).\square

\section*{Acknowledgments}
One of us (D.L.) is deeply indebted to the organizers of the MMRD 2000
Workshop in Leeds for stimulating
and warm atmosphere. Our particular thanks go to M.Semenov-Tian-Shansky
and E.  Sklyanin for numerous and illuminating discussions.

\bigskip\noindent
The research was partly supported by grants INTAS 97-1312;
RFBR 00-02-16477(S. Khar\-chev);
RFBR 00-02-16530 (D.Lebedev) and by grant 00-15-96557 for
Support of Scientific Schools.


\begin{thebibliography}{12}

\bibitem{Gu}M.Gutzwiller, {\it The quantum mechanical Toda lattice II},
Ann. of Phys., (1981), {\bf 133}, 304-331.

\bibitem{FM}H.Flaschka, D.McLaughlin, {\it Canonically conjugate
variables for the Korteweg-de Vries equation and the Toda lattice
with periodic boundary conditions}, Progr.Theor.Phys., (1976),
{\bf 55}, 438-456.

\bibitem{Skl1}E.Sklyanin, {\it The quantum Toda chain}, Lect.Notes
in Phys., (1985), {\bf 226}, 196-233.

\bibitem{Skl2}E.Sklyanin, {\it Separation of variables. New trends},
Progr.Theor.Phys.Suppl. (1995), {\bf 118}, 35-60;\ \
{\bf solv-int/9504001}.

\bibitem{KL1} S.Kharchev, D.Lebedev, {\it Integral representation for
the eigenfunctions of a quantum periodic Toda chain},
Lett.Math.Phys., (1999), {\bf 50}, 53-77; \ \ {\bf hep-th/9910265}.

\bibitem{Konst}B.Kostant, {\it Quantization and representation theory}
In: Representation theory of Lie Groups. Proc.SRC/LMS research
Symp., Oxford 1977, London Math. Soc. Lecture Notes, (1979),
{\bf 34}, 287-316.

\bibitem{Jac}H.Jacquet, {\it Fonctions de Whittaker associ\'ees aux
groupes de Chevalley}, Bull.Soc.Math. France, (1967), {\bf 95}, 243-309.

\bibitem{Sch}G.Schiffmann, {\it Int\'egrales d'entrelacement et
fonctions de Whittaker}, Bull.Soc.Math. France, (1971), {\bf 99}, 3-72.

\bibitem{Ha}M.Hashizume {\it Whittaker models for real reductive groups},
J.Math.Soc. Japan, (1979), {\bf 5}, 394-401.\\
{\it Whittaker functions on semisimple Lie groups}, Hiroshima Math.J.,
(1982), {\bf 12}, 259-293.

\bibitem{KL2} S.Kharchev, D.Lebedev, {\it Eigenfunctions of
$GL(N,\RR)$ Toda chain: The Mellin-Barnes representation},
Pis'ma v ZhETF, (2000), {\bf 71}, 338-343;\ \ {\bf hep-th/0004065}.

\bibitem{Skl3} E.Sklyanin, {\it Quantum Inverse Scattering Method.
Selected Topics}, in Quantum Group and Quantum Integrable Systems,
(Nankai Lectures in Mathematical Physics), ed. Mo-Lin Ge, Singapore:
World Scientific, (1992), 63-97; \ \ {\bf hep-th/9211111}.

\bibitem{GK}S.Gindikin, F.Karpelevich, {\it Plancherel measure of
Riemann symmetric spaces of nonpositive curvature}, Soviet Math. Dokl.,
{\bf 3}, (1962), 962-965.

\bibitem{CH}Harish-Chandra, {\it Spherical functions on a semisimple
Lie group. I}, Amer.J.Math., (1958), {\bf 80}, 241-310.

\bibitem{KLS} S.Kharchev, D.Lebedev, M.Semenov-Tian-Shansky, {\it Wave
functions for the open relativistic Toda chain}, in preparation.

\bibitem{PG}V.Pasquier, M.Gaudin, {\it The periodic Toda chain and a
matrix generalization of the Bessel function recursion relations},
J.Phys., (1992), {\bf A25}, 5243-5252.

\bibitem{ST}M.Semenov-Tian-Shansky. {\it Quantum Toda lattices. Spectral
theory and scattering.} Preprint LOMIR-3-84. Leningrad, 1984. 64p.\\
M.Semenov-Tian-Shansky, {\it Quantization of Open Toda Lattices.}
Encyclopaedia of Mathematical Sciences,
vol. 16. Dynamical Systems VII. Ch. 3. Springer Verlag, 1994, pp.226-259.


\end{thebibliography}
\end{document}